\documentstyle[preprint,aps,epsf]{revtex}
\begin{document}
\begin{center}
{\hfill{\normalsize{\bf MKPH-T-95-26}}\\[3cm]
\bf\Huge The Anomalous Decay $\eta\rightarrow\pi\pi\gamma\gamma$
\footnote{This work has been supported by the
Deutsche Forschungsgemeinschaft
(SFB 201)}}
\\[0.4cm]
G. Kn\"ochlein,
\footnote{Present address: Department of Physics and Astronomy, University
of Massachusetts, Amherst, MA 01003}
S. Scherer, D. Drechsel \\[0.4cm]
{\it Institut f\"ur Kernphysik, Johannes Gutenberg-Universit\"at}\\
{\it D-55099 Mainz, Germany}\\
\end{center}
\vspace{0.8cm}
With the commissioning of new powerful facilities with high
production rates the experimental investigation of eta decays
will become a precision tool for testing low-energy QCD in the
$SU\,(3)$ flavour sector.
The anticipated numbers of $10^8$ - $10^9$ observed etas
per year at DA$\Phi$NE, ITEP, CELSIUS and GRAAL will allow for
a test of chiral perturbation theory (ChPT, see Weinberg (1979) and
Gasser and Leutwyler (1984, 1985)) predictions up to
next-to-leading order or even beyond.  Experiments will include high-accuracy
measurements of the main decay modes $\eta\rightarrow\pi\pi\pi$
and $\eta\rightarrow\gamma\gamma$, as well as the detection of rare
decay modes predicted by ChPT. The mode $\eta\rightarrow\gamma\gamma$
involves the anomalous Wess-Zumino-Witten (WZW, Wess and Zumino
(1971), Witten (1983)) action and has been
thoroughly analysed in combination with the analogous $\pi^0$ and
$\eta'$ decays in theory as well as in experiment.
However, the WZW action also generates more complicated interactions,
which are of particular interest for the following reason:
at ${\cal{O}}(p^4)$ the WZW lagrangian completely covers the odd intrinsic
parity sector of ChPT and is the only
piece in the ${\cal{O}}(p^4)$ ChPT lagrangian which has actually
predictive power by itself, because its low-energy coefficient is
completely determined by the structure of the WZW functional and, in
contrast to the Gasser-Leutwyler lagrangian for the even intrinsic
parity sector, need
not be determined by means of phenomenological input. Hence, the
validity of the predictions derived from the WZW action is of
fundamental interest for the validity of ChPT and QCD itself. The
process $\eta\rightarrow\pi^+\pi^-\gamma\gamma$ has the remarkable
feature that it involves three different types of WZW interactions in
a consistent ${\cal{O}}(p^4)$ treatment according to Weinberg's power
counting (Weinberg (1979)). Such a calculation is a tree-level
calculation, because an odd number of
pseudoscalar mesons participates in the process
and, thus, one vertex in the Feynman diagrams contributing to the
process must have odd intrinsic parity. The invariant amplitude for
the processes $\eta\rightarrow\pi^+\pi^-\gamma\gamma$ and
$\eta\rightarrow\pi^0\pi^0\gamma\gamma$
is constructed from the standard ${\cal{L}}^{(2)}$ and the
anomalous ${\cal{L}}_{WZW}^{(4)}$ lagrangian,
\begin{eqnarray}
{\cal{L}} & = & {\cal{L}}^{(2)}+{\cal{L}}^{(4)}_{WZW} \nonumber\\
& = &
\frac{F_{\pi}^2}{4} {\mathrm{tr}} ((D^{\mu} U)^{\dagger} D_{\mu} U)
+ \frac{F_{\pi}^2 B_0}{2} {\mathrm{tr}}(M
(U + U^{\dagger}))\nonumber\\
& & +\frac{e N_c}{48 \pi^2}
\varepsilon^{\mu\nu\alpha\beta}
A_\mu {\mathrm{tr}}(Q\partial_\nu U\partial_\alpha U^\dagger\partial_\beta
U U^\dagger-Q\partial_\nu U^\dagger \partial_\alpha U \partial_\beta U^\dagger
U)\nonumber\\
& &
- \frac{i e^2 N_c}{24 \pi^2}
\varepsilon^{\mu\nu\alpha\beta}\partial_{\mu}
A_{\nu} A_{\alpha} {\mathrm{tr}}(Q^2(U \partial_{\beta} U^{\dagger}
+ \partial_{\beta} U^{\dagger} U )
-\frac{1}{2}Q U^{\dagger} Q \partial_{\beta} U + \frac{1}{2} Q U Q
\partial_{\beta} U^{\dagger}),
\end{eqnarray}
where only those terms of ${\cal{L}}_{WZW}^{(4)}$ have been retained
which give a contribution to the two processes under investigation.
The covariant derivative is defined as
$D_{\mu} U = \partial_{\mu} U + i e A_{\mu} \left[ Q, U \right]$,
where $Q={\mathrm{diag}} \left(2,-1,-1 \right)/3$ is the diagonal quark
charge matrix.
The current quark masses are contained in $M={\mathrm{diag}}(m_u,m_d,m_s)$,
and the constant $B_0$ is related to the quark condensate.
Since we want to include $\eta$-$\eta'$ mixing at a
phenomenological level, we assume nonet symmetry. Hence, the matrix
$U$ contains the pseudoscalar octet fields as well as the eta singlet
field.
We use the exponential representation of the pseudoscalar meson
fields,
$U={\mathrm{exp}}(i\Phi/F_{\Phi})$,
where $F_{\Phi}$ is the pseudoscalar meson
decay constant and the field matrix is defined as
\begin{equation}
\Phi =
\left(
\begin{array}{ccc}
\pi_0 + \frac{1}{\sqrt{3}} \eta_8 + \sqrt{\frac{2}{3}} \eta_0 & \sqrt{2}
\pi^+ & 0 \\
\sqrt{2} \pi^- & - \pi_0 + \frac{1}{\sqrt{3}} \eta_8 + \sqrt{\frac{2}{3}}
\eta_0 & 0 \\
0 & 0 & - \frac{2}{\sqrt{3}} \eta_8 + \sqrt{\frac{2}{3}} \eta_0
\end{array}
\right).
\end{equation}
The three types of Feynman diagrams for the charged
decay mode are
a diagram with a four-meson interaction from ${\cal{L}}^{(2)}$, a
$\pi^0$ or $\eta$ propagator and a one-meson-two-photon vertex from
${\cal{L}}_{WZW}^{(4)}$ (class 1 diagram), a diagram with a
three-meson-one-photon interaction from ${\cal{L}}_{WZW}^{(4)}$, where
another photon is emitted from a charged pion line (class 2 diagram)
and a contact diagram with a three-meson-two-photon interaction from
${\cal{L}}_{WZW}^{(4)}$. The class 1 diagrams are gauge invariant by
themselves, whereas for class 2 and 3 only the
sum of the diagrams is gauge invariant. It is interesting to note
that the matrix element for $\eta\rightarrow\pi^+\pi^-\gamma\gamma$ is
closely related to that of $\gamma\gamma\rightarrow\pi^+\pi^-\pi^0$
(see Bos (1994) for a treatment in ChPT). Thus, a measurement of the
decay could allow for a consistency check of the process with two
photons in the initial state.
The invariant amplitude for the neutral decay mode receives its only
contributions from class 1 diagrams.

Calculating the photon spin sum for the invariant matrix element
squared and integrating over four of the five independent
variables of the four-particle phase
space results in
the diphoton spectrum ${\mathrm{d}} \Gamma/{\mathrm{d}}z \,\,\,
(z=s_{\gamma}/m_{\eta}^2)$, where $s_{\gamma}$ is the invariant
c.m.\ energy squared of the diphoton system.
More details concerning
the calculation can be found in Bijnens {\it et al.} (1992) and
Kn\"ochlein {\it et al.} (1995). The diphoton spectrum of the charged
decay mode (Fig.1 (left)) clearly reflects a
bremsstrahlung singularity for $s_{\gamma}\rightarrow0$. Another
singularity of the invariant amplitude
due to the diagram with the propagating $\pi^0$ influences the
spectrum in a very narrow energy region around
$s_{\gamma}=m_{\pi^0}^2$ and could be removed by introducing the final
width of the $\pi^0$ in the propagator. In the rest of the spectrum
the influence of the diagrams with the propagating neutrals
is almost invisible. The partial width
$\Gamma(\eta\rightarrow\pi^+\pi^-\gamma\gamma)$ is calculated as a
function of an energy cut $\delta m_{brems}$ applied around
$s_{\gamma}=0$. Our result for $\delta m_{brems} = 195
\;{\mathrm{MeV}}$ is smaller than the experimental upper limits
(Price {\it et al.} (1967)) by about three orders of
magnitude. However, for $\delta m_{brems} \sim 30
\;{\mathrm{MeV}}$ our calculation yields a branching fraction of
approximately $10^{-4}$ which should be within the reach of the new
facilities.
The diphoton spectrum of the neutral decay mode (Fig. 1 (right)) is
clearly dominated by the $\pi^0$ pole diagram which is present due to
isospin and $G$ parity violation at the four-meson ${\cal{L}}^{(2)}$
vertex. The eta pole diagram, which is the only diagram present in the
limit of conserved isospin, plays a minor role in the dynamics of the
decay. The partial width for the neutral mode,
which is obtained by integration of the
diphoton spectrum and application of a reasonable energy cut around
$\sqrt{s_{\gamma}}=m_{\pi^0}$, is by more than two
orders of magnitude smaller than for the charged mode. Hence, it will
probably be extremely difficult to detect this mode.\\
[0.3cm]
REFERENCES\\[0.3cm]
Bijnens, J., G. Ecker and J. Gasser (1992), The
DA$\Phi$NE Physics Handbook, Vol. I (eds. L. Maiani, G. Pancheri and
N.
Paver), Servizio Documentazione dei Laboratori Nazionali di Frascati,
Frascati\\
Bos, J. W., Y. C. Lin and H. H. Shih (1994), Phys. Lett. {\bf{B337}} 152\\
Gasser, J. and H. Leutwyler (1984), Ann. Phys, {\bf{158}}, 142\\
Gasser, J. and H. Leutwyler (1985), Nucl. Phys. {\bf{B250}}, 465\\
Kn\"ochlein, G., S. Scherer and D. Drechsel (1995), Mainz university preprint
{\bf{MKPH-T-95-19}}\\
Price, L. R. and F. S. Crawford (1967), Phys. Rev. Lett. {\bf{18}}, 1207\\
Weinberg, S. (1979), Physica {\bf{96A}}, 327\\
Wess, J. and B. Zumino (1971), Phys. Lett. {\bf{B37}}, 95\\
Witten, E. (1983), Nucl. Phys. {\bf{B223}}, 422\\
\begin{figure}[h]
\centerline{
\epsfxsize=6.6cm
\epsfbox{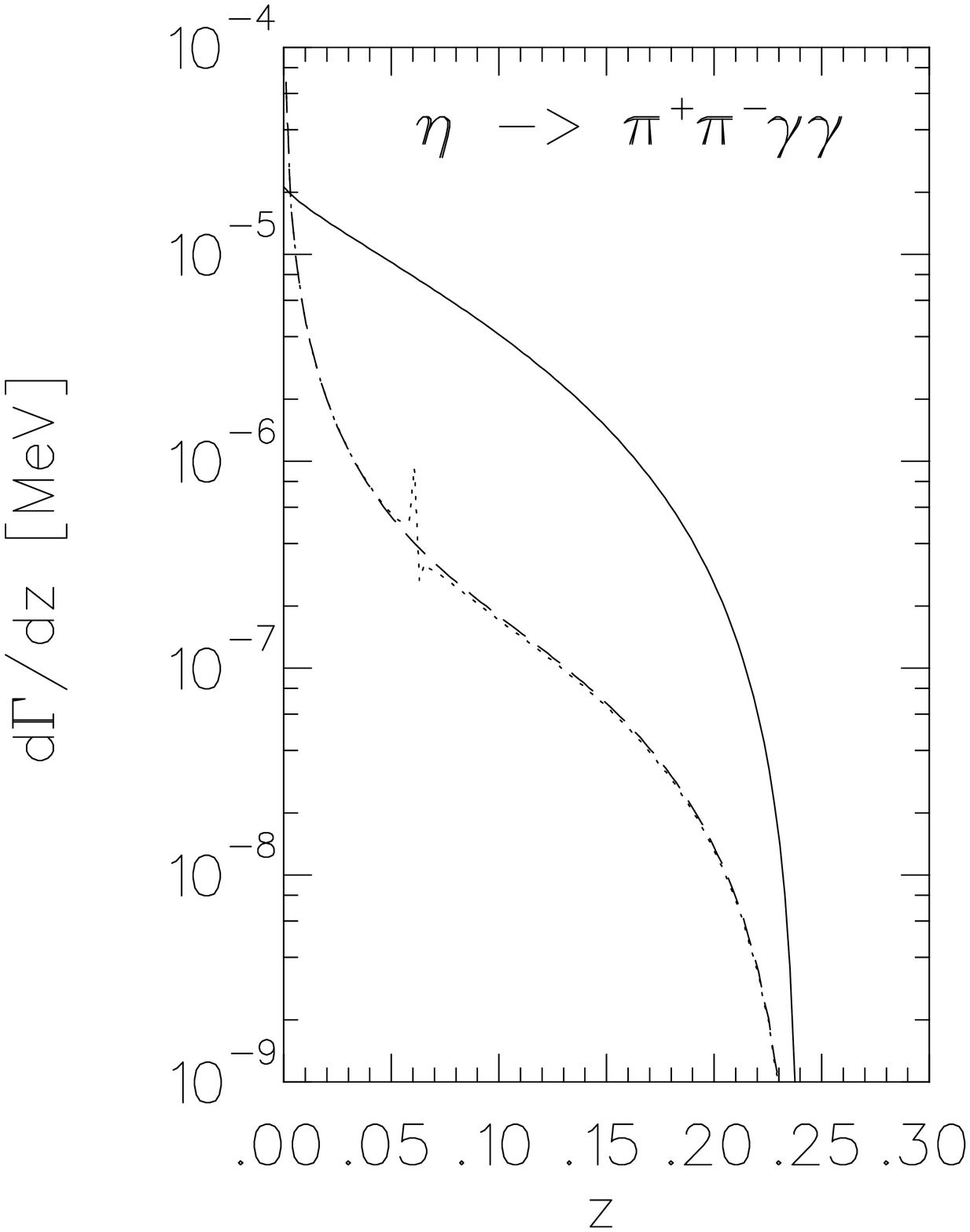}
\epsfxsize=6.92cm
\epsfbox{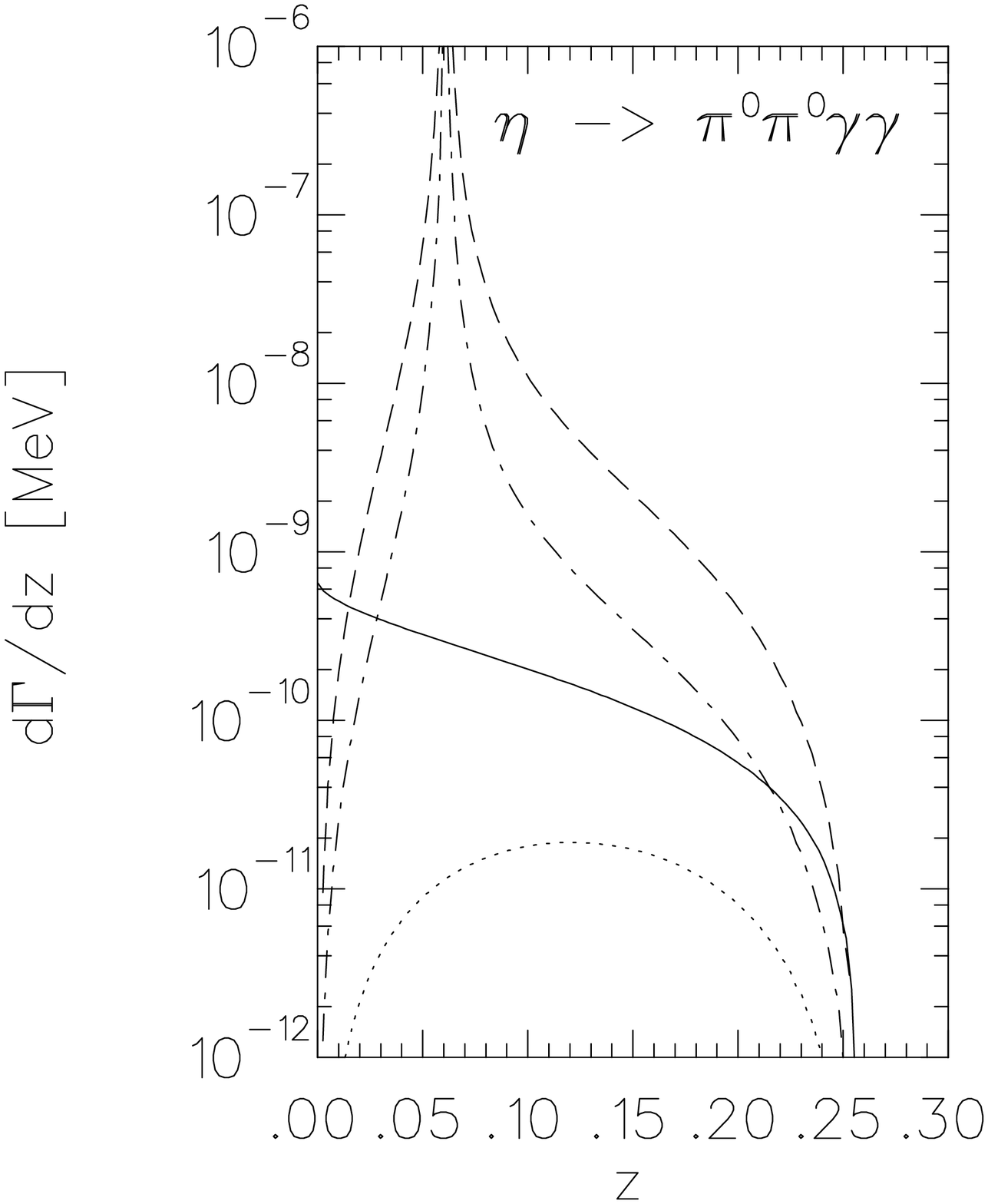}
}
\vspace{0.5cm}
\caption{
\small
Left figure: Diphoton energy spectrum ${\mathrm{d}} \Gamma/{\mathrm{d}}z \,\,\,
(z=s_{\gamma}/m_{\eta}^2)$
for the
decay $\eta\rightarrow\pi^+\pi^-\gamma\gamma$. The dotted
line is the full
calculation, the dashed line is calculated without class 1 diagrams
and the solid line is proportional to the phase space integral.\\
Right figure: Diphoton spectrum for the decay $\eta \rightarrow
\pi^0\pi^0\gamma\gamma$. The dash-dotted line is the
prediction of
chiral perturbation theory for $m_u = 5\,{\mathrm{MeV}}$ and
$m_d = 9\,{\mathrm{MeV}}$. The dotted line is the ChPT
prediction for $m_u = m_d = 7\,{\mathrm{MeV}}$.
The dashed line is a calculation, where we
determined the
strength of the $\eta \pi^0 \pi^0 \pi^0$ interaction from the decay
$\eta \rightarrow \pi^0\pi^0\pi^0$ and of the $\eta\eta'\pi^0\pi^0$
interaction
from the decay $\eta'\rightarrow\eta\pi^0\pi^0$, including also an
additional $\eta_8\eta_8\pi^0\pi^0$ interaction. The
solid line is
proportional to the phase space integral.
}
\end{figure}
\end{document}